\documentclass[letterpaper,twocolumn,preprintnumbers,secnumarabic,amsmath,amssymb,superscriptaddress]{revtex4}
\usepackage{amssymb,amsmath}
\usepackage{ulem} 
\usepackage{bm} 
\usepackage{booktabs} 
\usepackage{array}
\usepackage{latexsym}
\usepackage{graphicx}
\usepackage{color}
\usepackage{datetime}

\usepackage{tablefootnote}
\usepackage{multirow}

\usepackage{verbatim}
\usepackage{chngpage} 

\usepackage{psfrag}

\usepackage{mciteplus}

\usepackage[colorlinks=true,      linkcolor=blue,      urlcolor=blue,      
            filecolor=blue,      citecolor=blue,       pdfstartview=FitH,     
						pdfpagemode=UseNone,      bookmarksopen=true]{hyperref}  
\usepackage[all]{hypcap}     

\begin{document}
\title{A New Window into Black Holes}

 \author{Iosif Bena}

\author{Daniel R. Mayerson}
\affiliation{Institut de Physique Th\'eorique, Universit\'e Paris-Saclay, CNRS, CEA, Orme des Merisiers 91191, Gif-sur-Yvette CEDEX, France\\ \vspace*{1mm}{\rm \textsf{ iosif.bena@ipht.fr, daniel.mayerson@ipht.fr} }}

\begin{abstract}

We develop a formalism to compute the gravitational multipole moments and ratios of moments of non-extremal and of supersymmetric black holes in four dimensions, as well as of horizonless microstate geometries of the latter. For supersymmetric and for Kerr black holes many of these multipole moments vanish, and their dimensionless ratios are ill-defined. We present two methods to compute these dimensionless ratios, which for certain supersymmetric black holes agree spectacularly. We also compute these dimensionless ratios for the Kerr solution. Our methods allow us to calculate an infinite number of hitherto unknown parameters of Kerr black holes, giving us a new window into their physics. 
\end{abstract}

\maketitle

\newcommand{\IB}[1]{\textbf{\textit{\textcolor{magenta}{(IB: #1)}}}}
\newcommand{\drmdeb}[1]{\textbf{\textit{\textcolor{red}{(DRM: #1)}}}}
\newcommand{\todo}[1]{\textbf{\textcolor{red}{To do: #1}}}

\newcommand{\be}{\begin{equation}}
\newcommand{\ee}{\end{equation}}
\newcommand{\bea}{\begin{eqnarray}}
\newcommand{\eea}{\end{eqnarray}}
\newcommand{\bean}{\begin{eqnarray*}}
\newcommand{\eean}{\end{eqnarray*}}
\newcommand{\nn}{\nonumber}
\newcommand{\eps}{\epsilon}

 \def\indirect{\emph{indirect} }
 \def\Indirect{\emph{Indirect} }
 \def\direct{\emph{direct BPS} }
  \def\Direct{\emph{Direct BPS} }


\section{Introduction}
\label{sec:intro}
\vspace*{-2mm}

There is an extended literature that argues that in order for black hole evaporation to be consistent with quantum unitarity, there should exist a structure at the scale of the horizon of the black hole \cite{Mathur:2009hf, Almheiri:2012rt}. This structure, commonly referred as a fuzzball or firewall, has highly unusual properties in that its stiffness prevents its immediate collapse into the black hole. The only top-down construction of such structure is given by {\em black hole microstate geometries} \cite{Bena:2006kb,Bena:2015dpt,Bena:2017xbt,Heidmann:2017cxt,Bena:2017fvm,Heidmann:2019xrd,Bena:2020yii}, which are smooth horizonless solutions of String Theory that have the same mass and charge as a black hole but in which the horizon is replaced by a complicated structure of topologically-nontrivial bubbles wrapped by fluxes. 

Understanding how the physics of this structure differs from the physics of the black hole is of crucial importance, especially in the light of the recent observations of gravitational waves emitted when two black holes merge \cite{Abbott_2009}, and of future experiments that plan to explore Extreme Mass-Ratio Inspiral (EMRI) gravity waves \cite{Danzmann_1996} that should reveal very detailed information about horizon-scale physics. One important way in which microstate geometries differ from the black hole is in the higher multipole moments of the mass and angular momentum. Since EMRI gravitational waves are sensitive to many of these multipole moments and invariant ratios thereof, it is a crucial problem to calculate precisely these multipole moments for microstate geometries and to compare them to those of the corresponding black hole.

Most of the black hole microstate geometries that have been constructed so far correspond to extremal black holes and so do not allow us to make quantitative predictions that could be compared to what will measured from EMRI gravitational waves. However, one can use extremal black holes and their microstates to understand qualitatively the new black-hole physics that can be glimpsed from the gravitational multipoles of microstate geometries, much as one uses the ${\cal N}=4$ SYM quark-gluon plasma to understand qualitatively features of the quark-gluon plasma in the real world.

In this Letter, we compute the gravitational multipoles of generic non-extremal black holes in four dimensions, and of horizonless microstate geometries that have the same mass and charges as the supersymmetric extremal black hole. Because of its symmetry and lack of angular momentum, all the gravitational multipoles of the supersymmetric (BPS) black hole vanish, with the exception of the mass, $M_0$. However, for generic microstate geometries of this black hole all multipoles are finite. Furthermore, in the ``scaling limit'' in which the throat of the microstate geometries becomes very long, and they become more and more similar to the black hole, all their extra multipoles vanish and only $M_0$ survives. 

However, one can also consider ratios of multipole moments that stay finite in the scaling limit, such as the product of the angular momentum and the current quadrupole moment divided by the product of the mass and the mass octopole moment:
\be \label{eq:introratio}
{S_1 S_2 \over M_3 M_0}\,.
\ee

This, and many other multipole ratios, cannot be computed in the four-dimensional BPS black hole solution, where they are zero over zero. Hence, by computing these ratios in the scaling limit of various microstate geometries, we obtain a whole set of new quantities that characterize the BPS black hole. We will call this method of computing multipole ratios the \direct method.

There is another way in which  one can try to compute multipole ratios that are undefined in the black-hole geometry. One can deform the supersymmetric black hole into a nonextremal, rotating black hole, compute its multipole ratios and take back the supersymmetric, non-rotating limit. Similarly, we can use this \indirect method to compute multipole ratios that are undefined for the Kerr black hole: one can deform it into a general charged STU black hole, compute multipole ratios, and take back the charges to zero. In this way, we can associate well-defined multipole ratios with the Kerr black hole. These previously unknown  ratios provide new constraints for any model that parameterizes departures from the Kerr solution that may have an effect of gravitational waves.

For certain families of BPS black holes, the multipole ratios computed using these two methods are amazingly close. Given the very different types of solutions used for these calculation, this unexpected agreement is a strong indication that multipole ratios are intrinsic quantities that characterize black holes and their calculation gives us a new window into black-hole physics. 


We will first review the formalism to compute gravitational mass and current multipoles, and apply it to the most generic non-extremal STU black hole in four dimensions, and to several families of multicenter bubbling solutions that have the same charges and mass as the supersymmetric four-dimensional STU black holes. We will then use our two methods to compute multipole ratios for BPS black holes and discuss when these ratios agree. We also use the \indirect method to compute new hitherto unknown multipole ratios for the Kerr black hole.


\vspace*{-4mm}
\section{Gravitational Multipoles in Four Dimensions}
\vspace*{-2mm}

A coordinate-independent way to define multipole moments in a stationary, asymptotically flat four-dimensional spacetime was introduced by Thorne \cite{Thorne:1980ru} (and shown to be equivalent to the Geroch-Hansen formalism \cite{Geroch:1970cd, Hansen:1974zz, Gursel:equiv}). By using so-called ACMC-$N$ (asymptotically-Cartesian mass-centered to order $N$ \footnote{We will not distinguish between ``Cartesian'' $(x,y,z)$ coordinates and its spherical counterpart, $(r,\theta,\phi)$, related in the usual way.}) coordinates, one can read off the multipole moments from an asymptotic expansion of the metric. For stationary, axisymmetric spacetimes with Killing vectors $\partial_t, \partial_\phi$ (for which the $(l,m)$ multipoles are only non-zero for $m=0$), the asymptotic expansion of the metric components involving $t$ in an ACMC-$\infty$ coordinate system are given by:
\begin{equation}
\label{eq:themultipoleexpansion}
 \begin{aligned}
 g_{tt}  &= -1 + \frac{2M}{r}+ \sum_{l\geq 1}^{\infty}  \frac{2}{r^{l+1}} \left( M_l P_l + \sum_{l'<l} c^{(tt)}_{ll'} P_{l'} \right),\\
 g_{t\phi} &= -2r\sin^2\theta\left[ \sum_{l\geq 1}^{\infty} \frac{1}{r^{l+1}} \left( \frac{S_l}{l} P'_l  \!\!+\sum_{l'<l} c_{ll'}^{(t\phi)}  P'_{l'}\right) \right],
\end{aligned}
\end{equation}
where $P_l=P_l(\cos\theta)$ are Legendre polynomials. The terms that contain $c^{(ij)}_{ll'}$ correspond to non-physical ``harmonics'', and depend on the particular ACMC coordinates used. Note that for a given $l$, the only $c_{ll'}$ terms that may appear in the expansion have $l'<l$. The purely spatial metric components must also satisfy a similar expansion \cite{Thorne:1980ru,Cardoso:2016ryw}. Note that these coordinates remain asymptotically-Cartesian (AC-$\infty$) if one shifts the center of mass of the solution, but are only mass-centered when the mass dipole moment, $M_1$, is zero.

The coefficients $M_l, S_l$ are coordinate-independent; $M_l$ are the ``mass multipoles'' while $S_l$ are the angular-momentum or ``current multipoles'' of the metric. The most familiar ones are the mass $M = M_0$ and angular momentum $J = S_1$. \footnote{For a modern review of multipoles in general relativity and their application to astrophysical observables, see \cite{Cardoso:2016ryw} (especially Section 5.1 and the Appendix).}

\paragraph{Most general non-extremal STU black hole in four dimensions.}
This solution (\cite{Chow:2014cca} (Section 5.2)) depends on 11 parameters: the mass and rotation parameters, $m$ and $a$, as well as four electric/magnetic charge parameters, $\delta_I$/$\gamma_I$, for $I=0,\cdots,3$ and a NUT charge that we set to zero to avoid closed timelike curves. The mass and angular momentum of the black hole are:
\be \label{eq:genBHMJ} M = m\left( \mu_1 - \mu_2\frac{\nu_1}{\nu_2}\right), \quad  J = -ma\left(\frac{(\nu_1)^2}{\nu_2}+\nu_2\right)\ee
where $\mu_i,\nu_i$ are complicated combinations of the electric and magnetic parameters $\delta_I,\gamma_I$ (see equations (4.18), (4.19), (4.20), and (5.5) in \cite{Chow:2014cca} for the exact relations). The metric is given in \cite{Chow:2014cca} in terms of coordinates $(t,r,u,\phi)$. By performing the coordinate transformation $u\equiv -m\nu_1/\nu_2+a\cos\theta$ and then:
\be \label{eq:prolatetospher} r_S \sin\theta_S \equiv \sqrt{r^2 + a^2}\sin\theta, \qquad r_S \cos\theta_S  \equiv r\cos\theta,\ee
we obtain AC-$\infty$ coordinates $(t,r_S,\theta_S,\phi)$. We then shift the center of mass to obtain ACMC-$\infty$ coordinates, which allows us to read off the multipole moments of this black hole:
 \begin{align}
 \label{eq:app:generalBHmultipoles} M_{l} &= -\frac{i}{2} \left(-\frac{a}{M}\right)^l  Z \bar{Z}\left(Z^{l-1} - \bar{Z}^{l-1}\right),\\
S_{l} &= \frac{i}{2} \left(-\frac{a}{M}\right)^{l-1} \frac{J}{M}   \left(Z^{l} - \bar{Z}^{l}\right) \nn
\end{align}
where
 \be
 \label{eq:ZDdef} Z \equiv D - i M\,,~~~{\rm with}~~~ D \equiv m\left(\mu_2 + \frac{\nu_1}{\nu_2}\mu_1\right),
\ee

This general black hole reduces to the supersymmetric (static) D6-D2-D0 black hole when $a=0$ and $m=0$, and we can see that in this limit all of its dipole moments, except $M_0$, vanish. This makes perfect sense: four-dimensional supersymmetric black holes must have zero angular momentum and $SO(3)$ symmetry.

Furthermore, this solution reduces to the Kerr black hole when all charges vanish, which corresponds to $D=0$ and $J=Ma$. The multipoles in  (\ref{eq:app:generalBHmultipoles}) then become the known Kerr multipoles: $M_l+ i S_l = M(ia)^l$.

\paragraph{General Supersymmetric Bubbled Geometries.}

These horizonless solutions are smooth in five-dimensions and can have $\mathbb{R}^{4,1}$ or  $\mathbb{R}^{3,1} \times S^1$ asymptotics. The latter solutions have the same mass and charges as a four-dimensional supersymmetric black hole, and can be reduced to (singular) multicenter solutions of four-dimensional supergravity of the type constructed by Denef and Bates \cite{Bates:2003vx}. The four-dimensional metric is:
\be \label{eq:ds2multicenter} ds^2 = - (\mathcal{Q}(H))^{-{1\over 2}}(dt +  \omega)^2 + (\mathcal{Q}(H))^{1\over 2}\left( dx^2 +dy^2+dz^2\right).\ee
The solution is completely determined by 8 harmonic functions $H = (V, K^I, L_I, M)$ ($I=1,2,3$) on the flat $\mathbb{R}^3$ basis spanned by $(x,y,z)$ \cite{Gauntlett:2004qy, Bena:2005ni}. These harmonic functions are determined by the locations and residues of their poles, $\vec{r}_i$ ($i=1,\cdots, N$), which are commonly known as ``centers''. The coefficients $h_i$ together are the charges associated to the center $i$, collectively denoted by the charge vector $\Gamma^i$:
\be \Gamma^i = \left(v^i, k_1^i, k_2^i, k_3^i, l_1^i, l_2^i, l_3^i, m^i\right).\ee
The harmonic functions are then collectively given by:
\be H = h^0 + \sum_{i=1}^N \frac{\Gamma^i}{r_i},\ee
where $h^0$ are the moduli (values at infinity) associated to the harmonic functions and where $r_i\equiv |\vec{r}-\vec{r}_i|$ is the distance in $\mathbb{R}^3$ to the $i$'th center. The warp factors and rotation parameters of the five-dimensional solution are:
\begin{align}
\label{eq:quarticinvdef} \mathcal{Q}(H) & = Z_1 Z_2 Z_3 V - \mu^2 V^2, ~ Z_I = L_I + \frac{1}{2V} C_{IJK} K^J K^K,\\
 \mu &= M + \frac{1}{2V} K^I L_I + \frac{1}{6V^2}C_{IJK}K^I K^J K^K,
\end{align}
where for the STU model $C_{IJK} = |\epsilon_{IJK}|$.

We will only consider axisymmetric bubbling geometries, where all centers are on the $z$-axis (at positions $z_i$) and $\partial_\phi$ is a Killing vector.

Upon choosing canonical moduli corresponding to a D6-D2-D0 black hole,
\be\label{eq:pincermodulinom0}
(v^0, k^0_1,k^0_2,k^0_3, l^0_1,l^0_2,l^0_3,m^0) = (1,0,0,0,1,1,1,0),
\ee
we find the mass and current multipoles:
\begin{align}
\label{eq:multipolesMmod-simple} M_l &= \frac14 \sum_i \left[ v^i + l_1^i+l_2^i+l_3^i)\right] z_i^l,\\
\label{eq:multipolesSmod-simple} S_l &= \frac14 \sum_i \left[ - 2 m^i + k_1^i+k_2^i+k_3^i\right] z_i^l.
\end{align}
where we have implicitly shifted the origin of the $z$ coordinates to be at the center of mass of the solution. This is obtained by requiring that $M_1$ vanish, and ensures that the $(x,y,z)$ coordinates are ACMC-$\infty$.


\vspace*{-4mm}
\section{Multipoles of Horizonless BPS Microstate Geometries}
\vspace*{-2mm}
In this Letter we compute the mass and current multipoles for the three classes of horizonless microstate geometries that we construct in this section. These geometries are determined by the poles of $V,K_1,K_2$ and $K_3$ and by the moduli. To ensure that the solutions are smooth in five-dimensional supergravity, the $l_I^i$ and $m^i$ charges at the $i$'th center are given by:
\be \label{eq:lmcharges} l_I^i = -\frac12 C_{IJK} \frac{k^i_J k^i_K}{v^i}, \qquad m^i = \frac{1}{12} C_{IJK}\frac{k^i_I k^i_J k^i_K}{(v^i)^2},\ee

The moduli of the solutions we construct are
\be
(1,-2m_0,0,0,1,1,1,m_0)
\ee
where the value of $m_0$ is fixed by requiring no closed timelike curves and four-dimensional asymptotic flatness. To obtain a solution with canonical moduli \eqref{eq:pincermodulinom0} we further perform a gauge transformation:
\be \label{eq:gaugetransf} (K^1, L_I,M)\!\rightarrow\! (K^1, L_I,M)\!-\! c\,(-V,  C_{IJ1}  K^J, \frac{1}{2} L_1),\ee
with $c= 2\sum_i m^i/(1+\sum_i l_1^i)$. It is in this final gauge that we compute the multipole ratios of the solutions and we compare them to those of the corresponding black hole.

\paragraph{Solution $A$} 

The charges of this solution are those of the four-center scaling solution constructed in \cite{Heidmann:2017cxt}, but the asymptotics we consider is $\mathbb{R}^{3,1} \times S^1$. 
For simplicity of presentation, we give the $\tilde{k}_I^i$ parameters of the solution, which are related to the ${k}_I^i$ by
\be
\label{eq:kitildetot} k_I^i = \tilde{k}_I^i -v^i\tilde{k}^{\text{(tot)}}_I,  \quad  \tilde{k}^{\text{(tot)}}_I \equiv \sum_i \tilde{k}^i_I. 
\ee
The $v^i$ charges and $\tilde k_I^i$ parameters  are:
\begin{align}
 v^i &= \left( 1, 1, 12, -13 \right),\\
 \nn\tilde k^i_1 &= \left( -\frac{2087}{10000}, -\frac{678089}{1250}, \frac{55636379}{10000} + \hat{k}, \frac{3445309}{2000}\right),\\
 \nn\tilde k^i_2 &= \left( -\frac{491}{2500}, \frac{4712993}{1250}, \frac{30306499}{5000}, \frac{32175101}{5000} \right),\\
 \nn\tilde k^i_3 &= \left( \frac{1}{10000}, -\frac{49939}{10000}, -\frac{311181}{5000}, \frac{133657}{2000} \right),
\end{align}
and the $l_I^i, m^i$ charges are given by (\ref{eq:lmcharges}). 

 The solution has a  \emph{scaling limit} when $\hat{k} \approx -0.804597$. In this limit, the inter-center distances $r_{ij}$ collapse as $r_{ij}\rightarrow \epsilon\, r_{ij}$, and the throat of the solution becomes longer and longer, resembling more and more the black hole. However, the size of the bubbles at the cap of the solution remains the same  \cite{Bena:2006kb}, so the solution is smooth and horizonless for any $\epsilon>0$. In the $\epsilon\rightarrow 0$ limit, the solution appears to be virtually indistinguishable from the black hole, and all multipoles except the mass, $M_0$, vanish.
 
Note that both for this solution and the next ones, the charges $v^i$ and $\tilde k_I^i$ (or $k_I^i$) that we give are not integers. This is not a problem since in a scaling solution with four-dimensional asymptotics we can always multiply all the $v^i,k^i_I, l^i_I$ and $m^i$ charges by an overall coefficient, which results in a solution in which all these quantities are integers and all the multipole ratios remain the same. 

 \paragraph{Solution $B$}
This solution has the following  $v^i$ and $\tilde k_I^i$ charges: 
\begin{align}
 v^i &= \left(1, -156.96, 159, -2.04 \right),\\
 \nn \tilde k^i_1 &= \left(0.4951 + \hat{k}, -217.1, 166.6, -6.899\right),\\
\nn  \tilde k^i_2 &= \left(0.9053, -474.0, 461.6, -6.905\right),\\
 \nn \tilde k^i_3 &= \left(1.226, -68.79, 50.96, -0.6686\right),
\end{align}
and the scaling limit is reached when $ \hat{k} \approx 0.5354$.


\paragraph{Solution $C$}
The $v^i$ and $\tilde k_I^i$ charges are:
\begin{align}
 v^i &= \left(1.000, -1.896, 2.000, -0.1037 \right),\\
 \nn \tilde k^i_1 &= \left(0.7796 + \hat{k}, -20.99, 15.88, -7.329\right),\\
 \nn \tilde k^i_2 &= \left(0.4543, 2.452, -9.061, 0.1448\right),\\
 \nn \tilde k^i_3 &= \left(-0.09249, -5.241, 3.364, -0.2651\right),
\end{align}
and the scaling limit is reached when $ \hat{k} \approx -1.6122$.


\vspace*{-4mm}
 \section{A New Window into Black Holes}
\label{NewWindow}
\vspace*{-2mm}

As we have explained in the Introduction, for BPS black holes we can calculate multipole ratios both by taking the scaling limit of ratios calculated in multicenter BPS bubbling solutions (the \direct method), and by calculating these ratios in the general STU black hole and then taking the BPS limit (the \indirect method). As we will see below, for certain families of black holes, the multipole ratios computed using these two very different methods agree spectacularly. 

 On the other hand, one can also use our \indirect method to calculate multipole ratios that are ill-defined in the Kerr geometry. These are ratios containing one or more factors of $S_{2l}$ or $M_{2l+1}$ (for any $l$), which vanish for Kerr black holes. However, one can compute these ratios using our \indirect method by charging the Kerr black hole, evaluating the multipole ratios in the general STU black-hole background, and then taking back the Kerr limit. This procedure is well-defined as it gives a unique value for the multipole ratio in the Kerr limit (independent of how the charges are turned on and off).

In table \ref{tab:simplemultipolevals} we give several multipole ratios calculated using the \direct and \indirect methods for the BPS black holes corresponding to the A, B and C solutions, and using the \indirect method for the Kerr black hole.

\vspace*{-6mm}
\subsection{Multipole ratios for BPS black holes}
\vspace*{-4mm}

It is immediately clear that the multipoles computed using the two methods match extremely well for the BPS solutions $A$ and $B$ and rather poorly for solution $C$. We can quantify this by defining, for a given ratio $\mathcal{M}$, the ``mismatch parameter'' $\mathcal{E}^{(\mathcal{M})} \equiv \left| ( \mathcal{M}^{\text{(dir)}} - \mathcal{M}^{\text{(ind)}})/\mathcal{M}^{\text{(ind)}}\right|$,
where $\mathcal{M}^{\text{(dir)}},\mathcal{M}^{\text{(ind)}}$ are the ratios calculated using the two methods. When averaged over all entries of table \ref{tab:simplemultipolevals}, the three solutions have:
\be\label{eq:errvals} \mathcal{E}_{\text{ave}}(A) = 0.0451, \,\, \mathcal{E}_{\text{ave}}(B) = 0.000888, \,\, \mathcal{E}_{\text{ave}}(C) = 2.31 .\ee
  
\def\upmarg{4ex}
\def\downmarg{1.8ex}
\def\threedownmarg{5.4ex}
\newcommand{\prekerr}{$\protect\phantom{{}^{(*)}}$}
\newcommand{\postkerrdef}{$\protect\phantom{}^{(*)}$}
\newcommand{\postkerrundef}{$\protect\phantom{{}^{(*)}}$}
 \begin{table}[tb]\centering
\begin{tabular}{|c||c|c|c||c|c|c||c|}
\hline
  & \multicolumn{3}{c||}{\direct} &  \multicolumn{4}{c|}{\indirect} \\ 
 \emph{Ratio} & $A$ & $B$  & $C$&  $A$ & $B$ & $C$ & Kerr \\ \hline\hline
%
 \rule{0pt}{\upmarg} $\dfrac{M_2 S_3}{M_4 S_1}$ & 1.01 & 1.00  & 1.06 & 1& 1 &1 & \prekerr 1 \postkerrdef    \\[\downmarg]  \hline
  \rule{0pt}{\upmarg} $\dfrac{M_5 S_3}{M_3 S_5}$ & 1.18 & 1.20  & 1.08 & 1.20 & 1.20  & 1.32     & \prekerr 2 \postkerrundef \\[\downmarg] \hline
 \rule{0pt}{\upmarg} $\dfrac{M_3 M_3}{M_6 M_0}$ & -0.791 & -0.800  & 0.183 & -0.802 & -0.800  & -1.15    & \prekerr 0 \postkerrdef \\[\downmarg] \hline 
  \rule{0pt}{\upmarg} $\dfrac{M_3 M_3}{M_4 M_2}$ & 1.33 & 1.33  & 0.654 & 1.33 & 1.33  & 1.39    & \prekerr 0 \postkerrdef \\[\downmarg] \hline
  \rule{0pt}{\upmarg} $\dfrac{S_1 S_1}{M_2 M_0}$ & -28.6 & -135  & 1.06 & -22.6 & -136  & -6.20    & \prekerr -1 \postkerrdef \\[\downmarg]  \hline
   \rule{0pt}{\upmarg} $\dfrac{S_1 S_3}{M_2 M_2}$ & 87.1 & 405  & 3.27 & 67.8 & 407 & 16.1     & \prekerr -1 \postkerrdef\\[\downmarg] \hline
    \rule{0pt}{\upmarg} $\dfrac{M_3 S_2}{M_2 S_3}$ & 1.32 & 1.33  & 0.703 & 1.33 & 1.33 & 1.39  & \prekerr 0 \postkerrdef \\[\downmarg] \hline
 \rule{0pt}{\upmarg} $\dfrac{M_2 S_4}{M_4 S_2}$ & 0.676 & 0.667  & 1.29 & 0.666 & 0.667 &  0.615   &\prekerr 2 \postkerrundef \\[\downmarg] \hline
  \rule{0pt}{\upmarg} $\dfrac{M_2 S_2}{M_0 S_4}$ & -0.491 & -0.500  & 0.267 & -0.501 & -0.500 &  -0.627   & \prekerr $\dfrac12$ \postkerrundef \\[\downmarg] \hline
 \rule{0pt}{\upmarg} $\dfrac{M_4 S_4}{M_2 S_6}$ & 1.96 & 2.00  & 0.801 & 2.00 & 2.00 &  2.68   & \prekerr $\dfrac23$ \postkerrundef \\[\downmarg]  \hline
  \rule{0pt}{\upmarg} $\dfrac{S_3 S_2}{S_1 S_4}$ & 1.50 & 1.50  & 0.822 & 1.50 & 1.50 &  1.63   & \prekerr $\dfrac12$ \postkerrundef\\[\downmarg]  \hline
   \rule{0pt}{\upmarg} $\dfrac{M_2 S_2}{M_3 S_1}$ &  1.00 & 1.00 & 1.14 & 1 & 1 & 1   & \prekerr 1 \postkerrundef \\[\downmarg]  \hline
  \end{tabular}
\caption{Multipole ratios for the geometries $A,B,C$ computed using the two methods, and for Kerr computed using the \indirect method. The ratios marked with $(*)$
can be computed also in the Kerr geometry itself; all the other ratios are ill-defined in the Kerr geometry and can only be computed using our \indirect method.\vspace*{-.5cm}
}
\label{tab:simplemultipolevals}
\end{table}

Since we are working with four-dimensional BPS black holes whose five-dimensional uplift is a BMPV black hole in a Taub-NUT space, we can define an  \emph{entropy parameter} $\mathcal{H}$ which describes how close the BMPV black hole is to the cosmic censorship bound \cite{Heidmann:2017cxt}. In our four-dimensional solutions, this encodes how much entropy the black hole has compared to the entropy of a purely electric black hole with the same charges: 
\be 
\label{eq:entropypar} \mathcal{H} = \frac{\mathcal{Q}(Q_I,P_I)}{Q_1 Q_2 Q_3 Q_4},
\ee
where $Q_I$ are the D2,D2,D2 and D6 (electric) charges and $\mathcal{Q}(Q_I,P_I)$ is the quartic invariant (\ref{eq:quarticinvdef}) evaluated on the black hole charges, related to the entropy of the black hole as $S = \pi \sqrt{\mathcal{Q}(Q_I,P_I)}$ \footnote{Explicitly, we have \cite{Chow:2014cca}:
$4\mathcal{Q}(Q_I,P_I) =  4\prod_I Q_I + 4 \prod_I P_I + 2\sum_{I<J} Q_IQ_J P_I P_J - \sum_I Q_I^2P_I^2$.}.

Its values for the three solutions we consider is:
\be \label{eq:Hvals} \mathcal{H}(A) = 7.7\times 10^{-4},\,\, \mathcal{H}(B) = 7.9\times 10^{-6}, \,\, \mathcal{H}(C) = 0.055.\nonumber \ee
Comparing this to (\ref{eq:errvals}), it is clear that there is a correlation between small $\mathcal{H}$ and small ``mismatch'' $\mathcal{E}$. One can also analyze other microstate geometries to confirm this correlation \cite{ourupcoming}. We also checked that there is no correlation of (\ref{eq:errvals}) with any other quantity, such as the relative scale separation in distances between centers in the microstate geometries.


The spectacular agreement between the \direct and the \indirect methods of evaluating multipole ratios makes us confident that these multipole ratios are intrinsic characteristics of the black hole solutions, which we calculate in the Letter for the first time.

\begin{figure}[ht]\centering
 \includegraphics[width=0.4\textwidth]{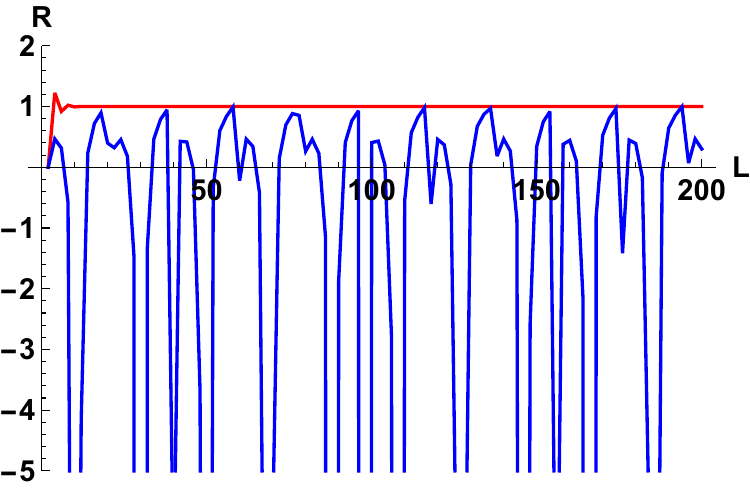}
 \caption{Multipole ratios $R(L) \equiv M_{L/2}M_{3L/2}/M_L^2$ (even $L$) in the geometry $C$, calculated using the \direct method (in red) and \indirect method (in blue).\vspace*{-.5cm}}
 \label{fig:oscillatingevidence}
\end{figure}

For the BPS black holes where the two methods do not give the same result, one can also wonder which method is more accurate. This will be discussed in detail \cite{ourupcoming}; however, it is easy to see that for certain multipole ratios, the \direct method gives a more reliable and sensible result. For example, in Fig. \ref{fig:oscillatingevidence} we show the multipole ratios $R(L) \equiv M_{L/2}M_{3L/2}/M_L^2$ calculated in the geometry $C$ for even $L$ between 2 and 200. The \direct method clearly gives a smooth, slowly varying result, whereas the \indirect method gives multipole ratios that are highly oscillating and discontinuous. 
\vspace*{-6mm}
 \subsection{Kerr Multipole Ratios}
\vspace*{-4mm}

An interesting result of our calculations is that some multipole ratios, such as those in the first and last rows of the table, are extremely close to one. These ratios are independent of the charges of the black hole and hence this value is universal, describing both supersymmetric and Kerr black holes, as well as everything else in between.

Furthermore, using our general result \eqref{eq:app:generalBHmultipoles} one can compute previously unknown ratios of vanishing multipole moments for the Kerr black hole, such as
\be
\label{eq:kerrratios} \frac{ M_2 S_l}{ M_{l+1} S_1} = 1~~~{\rm and} ~~~ \frac{ M_{l+2} S_{l} }{ M_{l} S_{l+2} } =   \frac{-1 + (-1)^l (2l+1)}{ 3 + (-1)^l (2l+1) }.
\ee
%
%
%
These and other similar ratios are independent of the rotation parameter $a$, so they also characterize the Schwarzschild black hole.
They in turn severely constraint the possible (small) deviations from Kerr multipoles that one may hope to measure using gravitational waves. Indeed, in the Kerr solution $M_{2n+1}=S_{2n} =0$ and our calculation predicts that, if there are small modifications of the Kerr black hole that make these multipoles finite, they must satisfy (with $\epsilon$ a small parameter):
\be \label{eq:constraint} M_{2n+1}=-a S_{2n} = n M a (-a^2)^n\, \epsilon,\ee

We believe that our results, and in particular the spectacular matching of BPS multipole ratios calculated using the \direct and \indirect methods, establishes that multipole ratios are an intrinsic feature of black holes. Furthermore, they can be used to place highly-nontrivial constraints on modifications of the Kerr multipole moments that produce a measurable effect on the gravitational waves emitted during black hole mergers.


\noindent
{\bf Acknowledgments}.  
We would like to thank Pierre Heidmann for sharing the unpublished scaling geometries B and C. We would also like to thank Vitor Cardoso and Bert Vercnocke for useful discussions. The  work  of  IB is  supported  by  the  ANR  grant  Black-dS-String  ANR-16-CE31-0004-01, by  the  John Templeton Foundation grant 61149, and the ERC Grants 787320-QBH Structure and 772408-Stringlandscape. DRM is supported by the ERC Starting Grant 679278 Emergent-BH.

\bibliography{bubble_multipoles}

\end{document}